\documentclass[11pt]{article}
\usepackage{amssymb}
\usepackage{epsfig}
\usepackage{apalike}
\usepackage{amssymb}
\def\sub{\atopwithdelims . .}
\begin{document}
\title{Instabilities in Attractor Networks with Fast Synaptic Fluctuations
and\linebreak Partial Updating of the Neurons Activity}
\author{J.J. Torres$^{+}$, J. Marro$^{+}$, J.M. Cortes$^{+\S \ast }$, and
B. Wemmenhove$^{\S }$
\\
\small
$^{+}$Institute \textquotedblleft Carlos I\textquotedblright\ for
Theoretical and Computational Physics, and \\
\small
Departamento de Electromagnetismo
y F\'{\i}sica de la Materia, 
\\
\small
University of Granada, 18071 Granada, Spain\\
\small
$^{\S }$Department of Biophysics and SNN, Radboud University, 
\\
\small
6525 EZ
Nijmegen, The Netherlands\\
\small
$^{\ast }$Institute for Adaptive and Neural Computation, 
School of
Informatics, 
\\
\small
University of Edinburgh, EH1 2QL, UK.}
\maketitle
\begin{abstract}
We present and study a probabilistic neural automaton in which
the fraction of simultaneously--updated neurons is a parameter, $\rho \in
\left( 0,1\right) .$ For small $\rho ,$ there is relaxation towards one of
the attractors and a great sensibility to external stimuli and, for $\rho
\geq \rho _{c},$ itinerancy among attractors. Tuning $\rho $ in this regime,
oscillations may abruptly change from regular to chaotic and vice versa,
which allows one to control the efficiency of the searching process. We
argue on the similarity of the model behavior with recent observations and
on the possible role of chaos in neurobiology.
\end{abstract}


\section{Introduction}

Attractor neural networks (ANN) are a paradigm for the property of
associative memory \cite{hopf1,hopf2}. Nevertheless, concerning practical
applications, and also when trying to mold the essence of actual systems,
the utility of ANN is severely limited, mainly by the fact that they can
only retrieve one memory at the time. In this note we show that such a
limitation may be systematically overcome by simply generalizing familiar
model situations. More specifically, we here extend some of our recent work
on ANN with fast pre--synaptic noise \cite{cortesNC,depre5,depre4}. The
result is a novel mathematically--tractable ANN whose activity eventually
describes heteroclinic paths among the attractors. This illustrates, in
particular, the possibility of a constructive role of chaos during searching
processes.

Our previous related studies essentially considered the same model system
but two different ways of updating it, namely, (\textit{i}) sequential and (%
\textit{ii}) parallel updating. Interesting enough, the ensuing behavior was
qualitatively, even dramatically different. That is, the main observation
was, respectively, (\textit{i}) a great enhancement of the system
sensibility to external stimuli as a consequence of rapid synaptic
fluctuations which simulate facilitation and/or depression \cite%
{cortesNC,depre5}, and (\textit{ii}) chaotic behavior while the system
spontaneously visited all the available attractors \cite{depre4}. Each of
these two regimes of behavior is to be associated with a different
functionality of an essential dynamic instability. Such an important
dependence on the updating process is rather unexpected. For instance, we
checked that it does not occur in a recent model \cite{depre1,depre1bis} which
is based on a different depression mechanism. This
situation motivated us to study in detail the changeover between (\textit{i}%
) and (\textit{ii}) as a modification of our previously proposed ANN \cite%
{cortesNC,depre4}. That is, we here present neural automata in which the
number or density $\rho $ of neurons that are updated at each time step is a
parameter. The resulting behavior as one modifies $\rho $ is varied and
intriguing. It leads us to argue on the possible relevance of our
observations to interpret neurobiological experiments.

\section{Definition of model}

Let the sets of neuron activities $\mathbf{\sigma }\equiv \left\{ \sigma
_{i}\right\} $ and synaptic weights $\mathbf{w\equiv }\left\{ w_{ij}\in 
\mathbb{R}
\right\} ,$ where $i,j=1,\ldots ,N,$ and assume a presynaptic current $%
h_{i}\left( \mathbf{\sigma },\mathbf{w}\right) $ on each neuron due to the
weighted action of the others. At each time unit, one updates the activity
of $n$ neurons, $1\leqslant n\leqslant N.$ This induces evolution in
discrete time, $t,$ of the state probability distribution according to 
\begin{equation}
P_{t+1}(\mathbf{\sigma })=\sum_{\mathbf{\sigma }^{^{\prime }}}R\left( 
\mathbf{\sigma }^{\prime }\mathbf{\rightarrow }\mathbf{\sigma }\right) P_{t}(%
\mathbf{\sigma }^{\prime }),
\end{equation}%
where the transition rate is a superposition: 
\begin{equation}
R\left( \mathbf{\sigma }\mathbf{\rightarrow }\mathbf{\sigma }^{\prime
}\right) =\sum_{\mathbf{x}}p_{n}(\mathbf{x})\prod_{\left\{ i|x_{i}=1\right\}
}\tilde{\varphi}_{n}\left( \sigma _{i}\rightarrow \sigma _{i}^{\prime
}\right) \prod_{\left\{ i|x_{i}=0\right\} }\delta _{\sigma _{i},\sigma
_{i}^{\prime }}.  \label{rate}
\end{equation}%
Here, $\tilde{\varphi}_{n}\left( \sigma _{i}\rightarrow \sigma _{i}^{\prime
}\right) \equiv \varphi \left( \sigma _{i}\rightarrow \sigma _{i}^{\prime
}\right) \left[ 1+\left( \delta _{\sigma _{i}^{\prime },-\sigma
_{i}}-1\right) \delta _{n,1}\right] $ and we denote $\mathbf{x}\equiv
\left\{ x_{i}=0,1\right\} $ an extra set of indexes which helps one in
selecting the desired subset of neurons. The above thus describes \textit{%
parallel} updating, as in familiar cellular automata \cite{droz},\textit{\ }%
for $n=N$ or, macroscopically, $\rho \equiv n/N\rightarrow 1,$ while
updating proceeds sequentially, as in kinetic Ising-like models \cite{marroB},
for $n=1$ or $\rho \rightarrow 0.$

We shall consider explicitly the simplest version of this model which
happens to be both interesting and mathematically tractable. First, we assume
binary neurons, so that $\sigma _{i}=\pm 1,$ which is known to be sufficient
in order to capture the essentials of cooperative processes \cite%
{depre1,marroB,binaryneurons}. The elementary rate $\varphi $ is an
arbitrary function of $\beta \sigma _{i}h_{i}$ (with $\beta $ an inverse
\textquotedblleft temperature\textquotedblright\ or stochasticity parameter)
which we assume to satisfy detailed balance. This property is not fulfilled
by the superposition (\ref{rate}) for $n>1,$ however. Consequently, the
resulting steady states are generally out of equilibrium, which is more
realistic in practice than thermodynamic equilibrium \cite{marroB}. On the
other hand, we shall only illustrate the case in which the $n$ neurons are
chosen at random out from the set of $N,$ so that one has $p_{n}\left( 
\mathbf{x}\right) =\allowbreak \left (N\sub n \right )^{-1}\delta \left(
\sum_{i}x_{i}-n\right) $ in (\ref{rate}). For the sake of simplicity, we
also need to assume that the currents are such that $h_{i}\left( \mathbf{%
\sigma },\mathbf{w}\right) =h\left[ \mathbf{\pi }\left( \mathbf{\sigma }%
\right) ,\mathbf{\xi }_{i}\right] ,$ where $\mathbf{\xi }_{i}\equiv \left\{
\xi _{i}^{\mu }=\pm 1;\mu =1,\ldots ,M\right\} $ are some given, \textit{%
stored patterns} (realizations of the set of activities) and $\mathbf{\pi }%
\equiv \left\{ \pi ^{\mu }\left( \mathbf{\sigma }\right) \right\} .$ Here, $%
\pi ^{\mu }\left( \mathbf{\sigma }\right) =N^{-1}\sum_{i}\xi _{i}^{\mu
}\sigma _{i}$ measures the \textit{overlap} between the current state and
pattern $\mu .$ For $N\rightarrow \infty $ and finite $M,$ i.e., in the
limit $\alpha \equiv M/N\rightarrow 0$ (which is not the interesting case, but may serve 
first for illustrative purposes) the resulting time equation under these conditions is $
\pi _{t+1}^{\mu }\left( \mathbf{\sigma }\right) =\rho
N^{-1}\sum\nolimits_{i}\xi _{i}^{\mu }\tanh \left( h_{i}^{t}\right) +\left(
1-\rho \right) \pi _{t}^{\mu }\left( \mathbf{\sigma }\right) ,  \label{mt}
$
where $h_{i}^{t}\equiv \beta h_{i}\left[ \mathbf{\pi }_{t}\left( \mathbf{%
\sigma }\right) ,\mathbf{\xi }_{i}\right] ,$ for any $\mu .$ The above result is general and valid for any type of patterns. It is to be noticed that the sum over $i$ in this map can be replaced by an average over the distribution 
of patterns $p(\xi^{\mu}_i).$ This permits a simple derivation of mean-field dynamical equations for the overlaps, at least for finite $M.$ Note also that Monte Carlo simulations do not require restriction concerning the nature of the stored patterns.

The above allows for different relations between the currents $h_{i}$ and
the weights $w_{ij},$ and between these and other system properties. The
simplest realization corresponds to the Hopfield case \cite{hopf1} which
follows from the map above for $\rho \rightarrow 0$ and currents given by $%
h_{i}\left( \mathbf{\sigma },\mathbf{w}\right) =\sum_{j\neq i}w_{ij}\sigma
_{j}$ with the weights fixed according to the Hebb prescription, namely, $%
w_{ij}=N^{-1}\sum_{\mu }\xi _{i}^{\mu }\xi _{j}^{\mu }.$ The symmetry $%
w_{ij}=w_{ji}$ then assures $P_{t\rightarrow \infty }\left( \mathbf{\sigma }%
\right) \propto \exp \left( \beta \sum_{i}h_{i}\sigma _{i}\right) $ and, for
high enough $\beta ,$ the \textit{stored }patterns $\mathbf{\xi }$ are
attractors of dynamics \cite{hopf2}. We checked that, in agreement with some
indications \cite{herzPRE}, the Hopfield--Hebb network exhibits
associative memory for any $\rho >0.$ However, the situation is more complex,
e.g., it depends on $\rho ,$ as one goes beyond Hopfield--Hebb, as we show
in the next section.

It is well documented that transmission of information and computations in
the brain are correlated with activity--induced fast fluctuations of
synapses, i.e., our $w_{ij}$'s \cite{noise,dobrunz,abb}. More specifically, it has
been observed that there is some efficacy lost after heavy work, so that
synapses suffer from \textit{depression}; it is claimed that repeated
activation decreases the neurotransmitter release which depresses the
synaptic response \cite{depre0,mdepre1,abb0,depre00,mdepre3}. The
consequences of this have already been analyzed in various contexts \cite%
{depre1,mdepre3,bibit,cortesNC,depre4,depre5}, and a main general conclusion from
these studies is that depression importantly affects a network performance reducing, in particular, the stability of the attractors.
Motivated by these facts, we shall adopt here the Hopfield currents and the
following prescription for the synaptic weights:%
\begin{equation}
w_{ij}=\left[ 1-\left( 1-\Phi \right) q\left( \mathbf{\pi }\right) \right]
N^{-1}\sum\nolimits_{\mu }\xi _{i}^{\mu }\xi _{j}^{\mu },  \label{wij}
\end{equation}%
where $q\left( \mathbf{\pi }\right) \equiv \frac{1}{1+\alpha }\sum_{\mu }\pi
^{\mu }\left( \mathbf{\sigma }\right) ^{2}.$ Note here that, in addition of static quenched disorder as in the standard Hopfield model, the weights (\ref{wij}) include a time dependence through the overlap vector $\mathbf{\pi }$ which is a measure of the network firing activity. These weights, which reduce to the Hebb prescription for $\Phi =1,$
amount to assume short--term fluctuations which change synapses by a factor $%
\Phi $ on the average with a probability $q\left( \mathbf{\pi }\right)$. Therefore, any positive $\Phi < 1$ simulates synaptic depression if $q\left( \mathbf{\pi }\right)$ is large. This is in agreement with the fact that, the greater $\pi$ is, more activity will in the average arrive to a particular postsynaptic neuron $i$ in the network and, therefore, this neuron will be more depressed. Although the magnitude $q\left( \mathbf{\pi }\right)$ involves a sum over all stored patterns, this will only affect neurons that are active in a particular pattern for not too high correlated patterns. More details concerning these assertions are in \cite{cortesNC,depre4}. 

Our setting here is rather close to the one in previous treatments of depressing synapses in a cooperative environment. As a matter of  fact, one may show after some simple algebra
that the model in \cite{depre1,storage,depre1bis} corresponds to certain choices of $%
\Phi $ and $q\left( \mathbf{\pi }\right) $ in (\ref{wij}) concerning steady states. For instance, a possible choice for $M=1$ and $\rho=1$ is $\Phi=1-\gamma/\gamma_0$ and $q (\pi)=\frac{\gamma_{0} [\gamma (1-\pi^2)+4]}{\gamma^2(1-\pi^2)+4\gamma+4}$ where $\gamma$ is the depression parameter defined in \cite{storage} and $\gamma_0$ is the value for that parameter at which $\Phi=0.$ This type of nonlinearity in $q\left(\pi\right),$ however, induces less susceptibility than the choice we are using here (see next section). 

For the sake of completeness, we shall be concerned in this paper with both positive and negative values of $\Phi .$
A result is that the behavior we are looking for ensues in any of these cases (but only for certain values of $\Phi$).

\section{Some main results}

\begin{figure}[ht!]
\begin{center}
\psfig{file=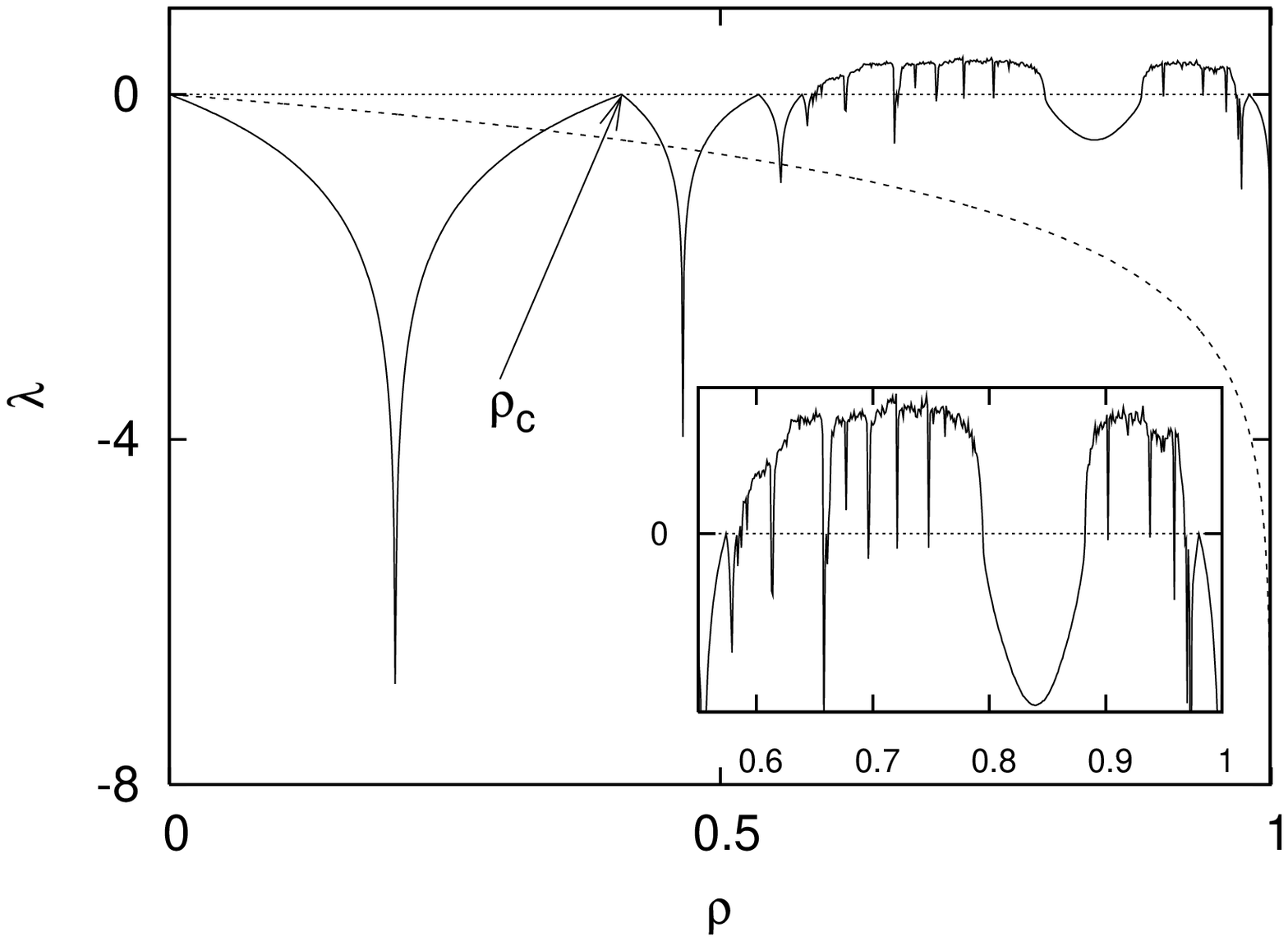, width=12cm}
\end{center}
\caption{Torres et al.}
\label{fig1}
\end{figure}

In the limite $N\rightarrow \infty$ the (nonequilibrium) stationary state follows from the map for $M=1$ as $%
\pi _{\infty }=F\left( \pi _{\infty };\rho ,\Phi \right) ,$ and local
stability requires that $\left\vert \partial F/\partial \pi \right\vert <1;$ 
$F\left( \pi ;\rho ,\Phi \right) \equiv \rho \tanh \left\{ \beta \pi \left[
1-\left( 1-\Phi \right) \pi ^{2}\right] \right\} \allowbreak +\left( 1-\rho
\right) \pi .$ The fixed point is therefore independent of $\rho ,$ but
stability demands that $\rho <\rho _{c}$ with%
\begin{equation}
\rho _{c}=2\left\{ 3\beta \pi _{\infty }^{2}\left[ \left( {\frac{4}{3}-}\Phi
\right) -\left( 1-\Phi \right) \pi _{\infty }^{2}\right] -\beta +1\right\}
^{-1}  \label{roc}
\end{equation}%
(a condition that cannot be fulfilled in the Hopfield, $\Phi =1$ case). As
Fig. \ref{fig1} shows, $\rho =\rho _{c}$ marks the period-doubling route to chaos in the
saddle--point map. This behavior is confirmed 
numerically for $M\gg 1$ stored arbitrary patterns, as shown numerically below.

\begin{figure}[ht!]
\begin{center}
\psfig{file=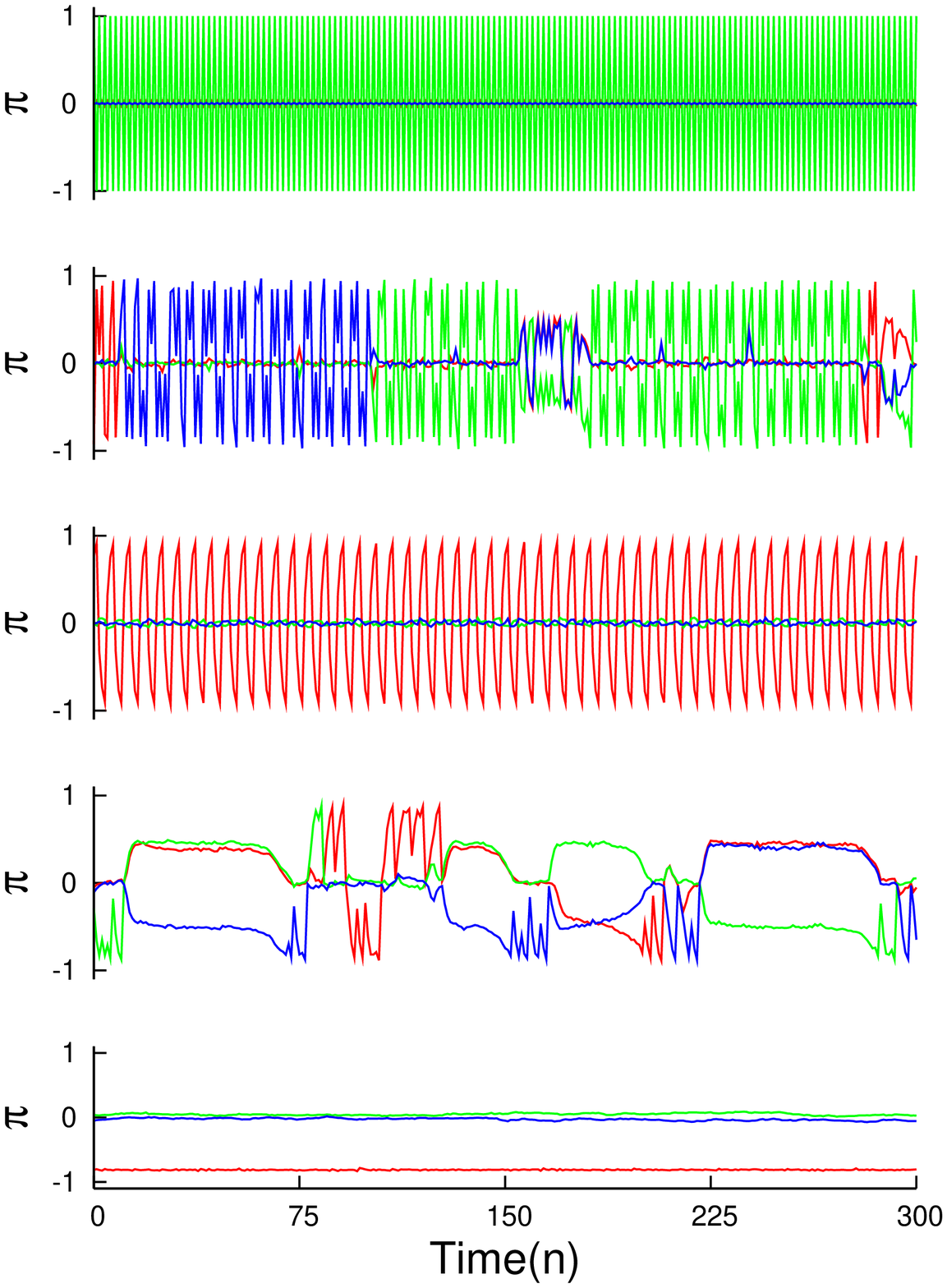,width=12.cm}
\end{center}
\caption{Torres et al.}
\label{fig2}
\end{figure}

Fig. \ref{fig2} shows some typical \textit{stationary} Monte Carlo runs,
i.e., from bottom to top: (\textit{a}) convergence towards one attractor
---in fact, one of the \textit{antipatterns}, namely, the negative of one of
the given patterns--- for small $\rho ;$ (\textit{b}) fully irregular
behavior with positive Lyapunov exponent for $\rho >\rho _{c};$ (\textit{c})
regular oscillation between one attractor and its negative for $\rho >\rho
_{c};$ (\textit{d}) onset of chaos as $\rho $ is further increased; and (%
\textit{e}) rapid and ordered periodic oscillations between one pattern and
its antipattern when, finally, all the neurons are active. The cases (%
\textit{b}) and (\textit{d}) are examples of instability--induced switching
phenomena, in which the system activity chaotically \textit{visits}
different attractors by describing heteroclinic paths and remaining
different time intervals in the neighborhood of each attractor. This kind of
behavior was previously observed for $\rho \rightarrow 1$ at certain values
of $\Phi $ \cite{cortesNC}. The interesting new facts are that this requires
a minimum of synchronized neurons, that this minimum ---as
well as many other details--- depend on $\Phi ,$ and that, as we show in the caption of figure \ref{fig1},
varying $\rho $ above $\rho _{c}\left( \Phi \right) $ seems to induce
further intriguing qualitative changes.

\begin{figure}[ht!]
\begin{center}
\psfig{file=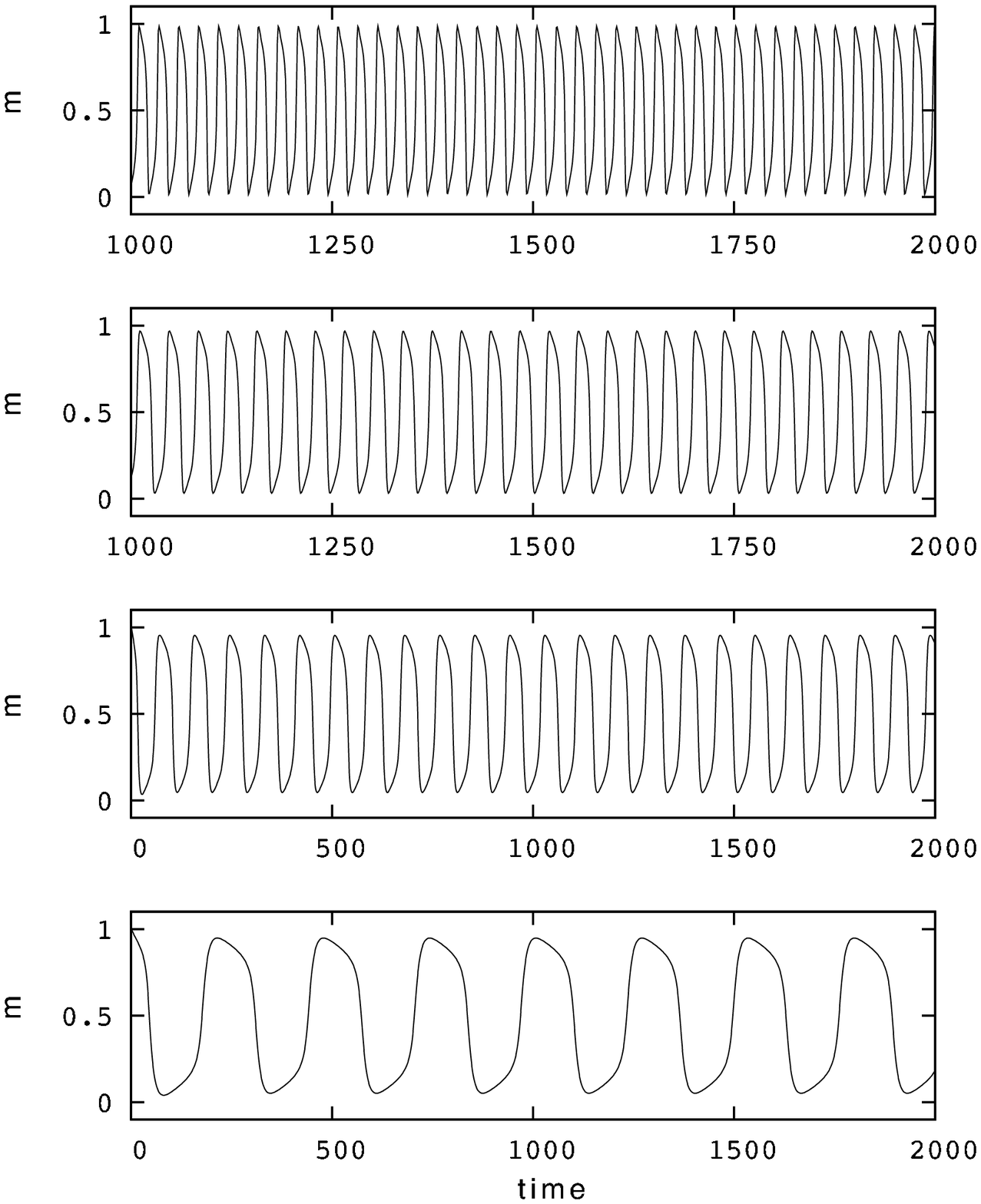, width=10cm}
\end{center}
\caption{Torres et al.}
\label{fig3}
\end{figure}

\begin{figure}[ht!]
\begin{center}
\hspace{-0.5cm}\psfig{file=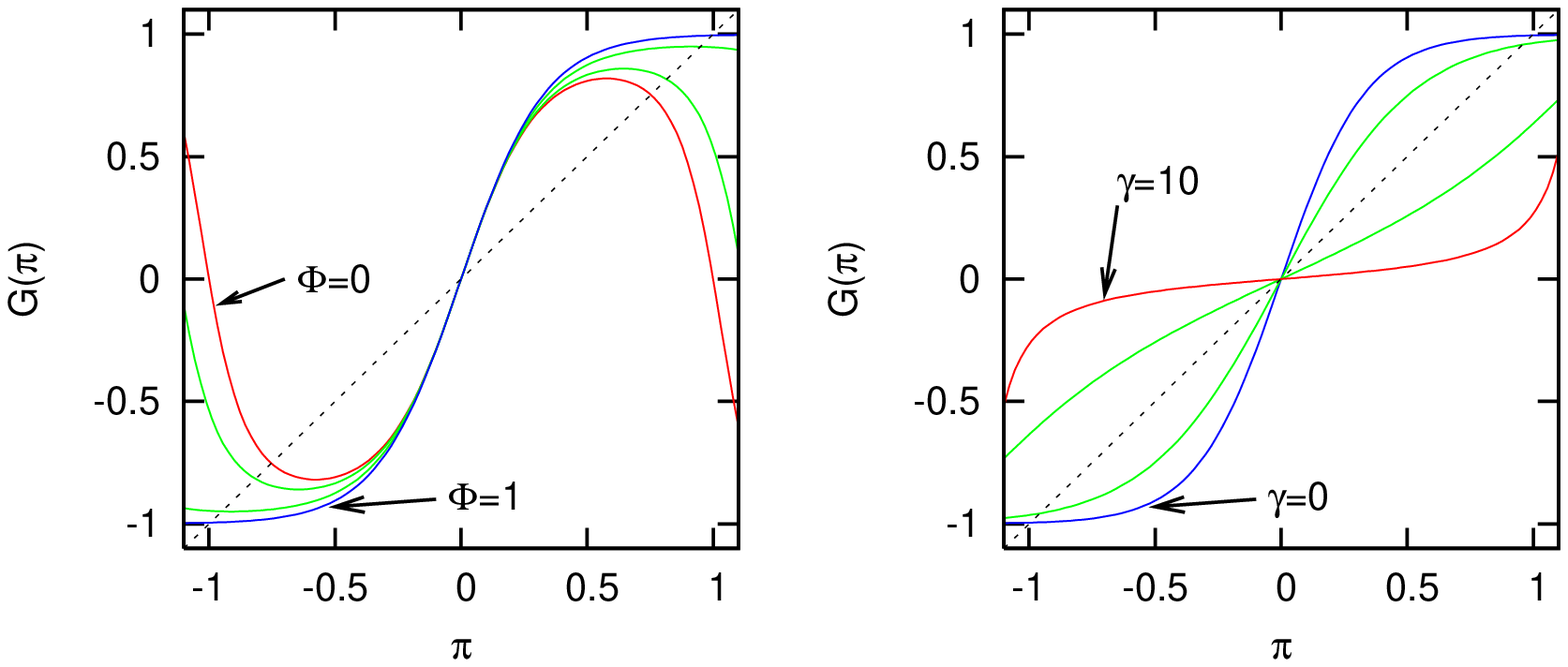, width=15cm}
\end{center}
\caption{Torres et al.}
\label{fig4}
\end{figure}

It is also to be remarked that chaotic switching or itinerancy requires that
the system is in a specially susceptible state first described in \cite%
{cortesNC,depre5}. This is accomplished in the present case by means of the
activity--dependent fast noise modelled in (\ref{wij}). One should expect
that variations of this assumption on the weights may result in an
equivalent susceptible state. As a matter of fact, we found that changing
the sign of $\Phi $ does not affect our main observations. However, the case 
$\Phi =1,$ in which the weights are fixed, does not exhibit interesting behavior, 
and $\rho $ turns then into an irrelevant parameter. On the
other hand, the model in \cite{depre1,storage,depre1bis} does not seem to
involve sufficient susceptibility for the purpose (see figure \ref{fig3}), in spite of the fact that
it includes an activity--dependent depression mechanism. The explanation is the following. Assuming that the dynamics can be writen as $\pi_{t+1}=G(\pi_t),$ the gain function $G(\pi)$ in the model in \cite{storage} is a nonlinear one which behaves monotonically for all values of the depression parameter. In our case, however, a non-monotonic type of gain function occurs for some values of $\Phi$ and $\rho$ (see comparison in figure \ref{fig4}). This has been reported to be important to originate a chaotic dynamics among the attractors \cite{dominguez97,caroppo99}. 

Monitoring activity trajectories as one varies $\rho $ in the case of
several stored patterns provides the following qualitative picture for
arbitrary patterns. As far as $\rho <\rho _{c},$ the activity remains
wandering around one of the patterns. The pattern selected depends on the
initial condition, and the trajectory visits a neighborhood of it whose
volume increases slightly with $\rho .$ The trajectory seems to tend to
densely fill this volume with time. Increasing $\rho,$ however,
the system may escape from the initially chosen pattern and, eventually, will tend
to visit all the patterns. In addition, one observes that the trajectory is
rather structured. That is, there are many jumps between the more correlated
patterns but only very few to the less correlated ones if the system is
close to the edge of chaos, and the system attention to all the patterns tends to
be balanced as $\rho $ is increased within a chaotic
window. Increasing $\rho $ further, the network surpasses equiprobability of
patterns and, eventually, abandons the chaotic regime to fall into a limit
cycle, where periodically oscillates between a pattern and its antipattern.
This confirms and details the behavior shown in figure \ref{fig1}. 

This behavior, which is clearly observed in Monte Carlo simulations, can also be obtained under a mean field theory.
Assume, for instance, random patterns with $p(\xi^\nu_i)=\frac{1+a}{2}\delta(\xi^\mu_i -1)+  \frac{1-a}{2}\delta(\xi^\mu_i +1)$ where $\langle \xi^{\nu}\rangle=a,$ even for $0<|a|\ll 1.$ In the simplest case of two patterns this mean field dynamics is determined by
\begin{equation}
\begin{array}{l}
\displaystyle \pi^1_{t+1}=\rho\;\frac{1+a^2}{2}\tanh [B(\pi_t)(\pi^1_t\!+\!\pi^2_t)]+\rho\;\frac{1-a^2}{2}\tanh [B(\pi_t) (\pi^1_t\!-\!\pi^2_t)]
\\
\\
\quad\quad\quad+\,(1-\rho)\;\pi^1_t
\\
\\
\displaystyle \pi^2_{t+1}=\rho\;\frac{1+a^2}{2}\tanh [ B(\pi_t)(\pi^1_t\!+\!\pi^2_t)]-\rho\;\frac{1-a^2}{2}\tanh [B(\pi_t)(\pi^1_t\!-\!\pi^2_t)]
\\
\\
\quad\quad\quad+\,(1-\rho)\;\pi^2_t,
\end{array}
\label{system2}
\end{equation} 
where $ B(\pi)\equiv\beta[1-(1-\Phi)q(\pi)].$ It may be noticed that only in the non-interesting case of orthogonal patterns, namely $a=0,$ the mean field dynamics (\ref{system2}) gives chaotic switching between a particular pattern and its antipattern but not between different patterns. Otherwise, the situation is of chaotic switching among the stored patterns.
\section{Discussion}

This paper deals with ANN in which the density $\rho $ of neurons that are
updated at each time step is a parameter, so that the limit $\rho
\rightarrow 0$ (1) corresponds to sequential (parallel) updating. Our main
motivation is that previous studies of ANN in these limits revealed
qualitatively different behavior, and that analysis in which the number of
updated neurons is systematically varied are rare in the literature, e.g.,
\cite{herzPRE}. It is worth to remark also that there are several arguments which
suggest studying changes with $\rho .$ One is simply the suspicion, born
outside biology, that a network could perhaps like to maintain inert some of
the nodes during operation, and not necessarily for economy but in order to
gain efficiency. As a matter of fact, as one may get convinced by oneself by
looking at our expressions for the currents $h_{i},$ hushing some of the
nodes may be equivalent to modifying the wiring topology, and this is
recognized as a method to enhance a network efficiency \cite{ournet}. More
specifically within biology, one may notice that assuming cells that are
stimulated only in the presence of a neuromodulator such as dopamine, $\rho $
could stand for the fraction of neurons modulated each cycle. There is no
input on the other $1-\rho ,$ so that information from the previous state is
maintained, which was argued to be a basis for working memories \cite%
{micro,workmem}. On the other hand, varying $\rho $ may also be relevant to
simulate various situations of persistent activity \cite{pers}, the observed
variability of the neurons threshold \cite{threshold}, and the possible
existence of \textit{silent neurons} \cite{silent,silent2}, for instance.

The fact is that varying $\rho $ in our model turns out to be very
intriguing. However, $\rho $ is relevant only if the network is \textit{%
susceptible}. Such a condition occurs in our case as a consequence of
activity--dependent \textit{fast synaptic} noise as modelled in (\ref{wij}).
The parameter $\rho $ is irrelevant in other cases as, in particular, for
the model in \cite{depre1,depre1bis} which is based on the depression mechanism introduced in \cite{depre0}, and also when the synaptic weights are fixed, even heterogeneously as in a Hopfield--Hebb network. On the contrary,
the model here exhibits kind of \textit{dynamic} association, namely, the
activity either goes to one attractor or else, for large enough $\rho,$
visits possible attractors. The visits may abruptly become chaotic. Besides
synchronization of a minimum of neurons, this requires careful tuning of $%
\rho .$ As a matter of fact, as shown by equation (\ref{roc}) and figure \ref%
{fig1}, a complex situation makes it difficult to predict the result for
slight changes of $\rho .$%

\begin{figure}[ht!]
\begin{center}
\psfig{file=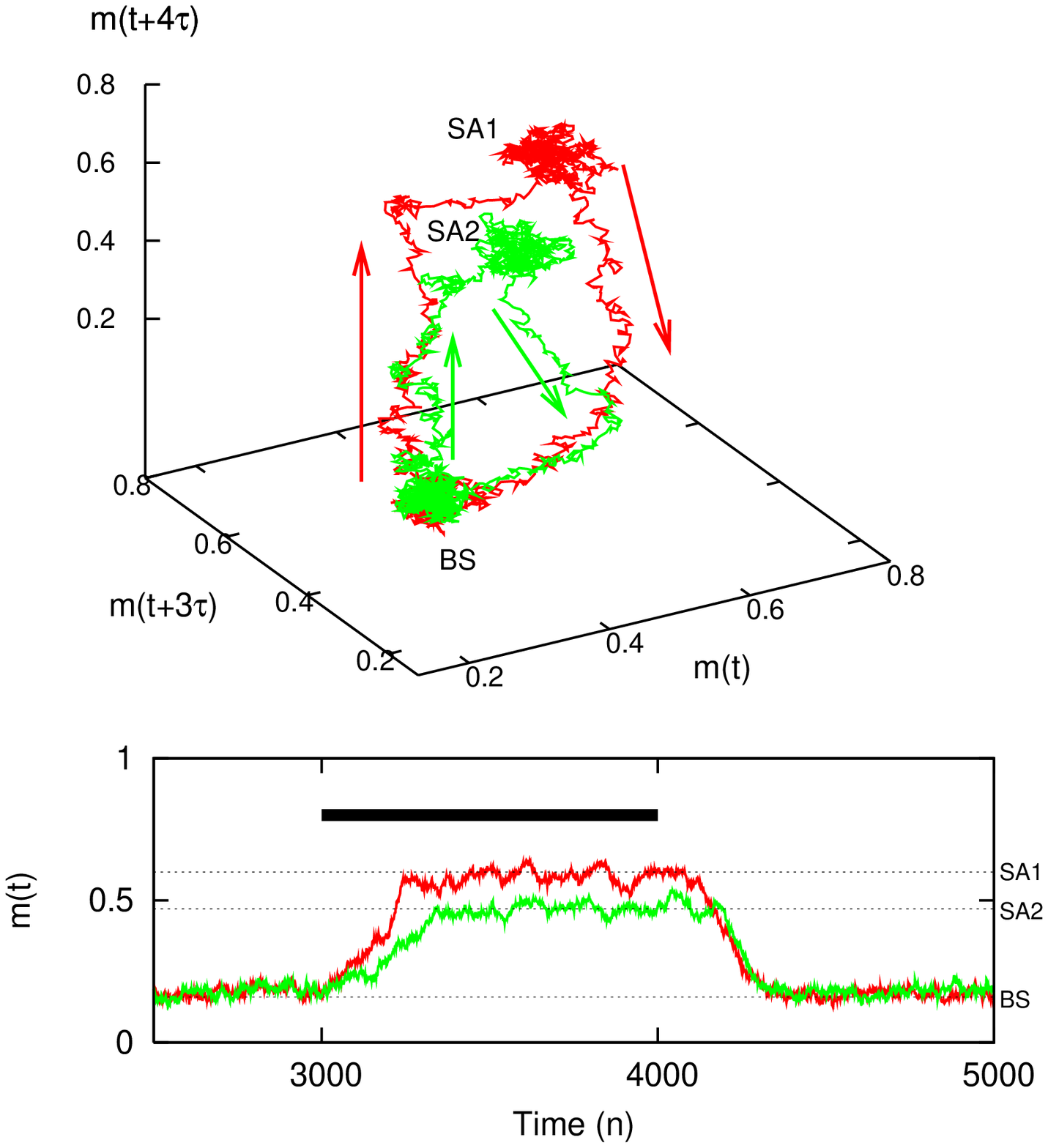,width=11.cm}
\end{center}
\caption{Torres et al.}
\label{fig5}
\end{figure}

Another interesting feature of our model is illustrated in figure \ref{fig5}.
This shows time series of the mean firing rate, $m=\frac{1}{2N}%
\sum_{i}\left( 1+\sigma _{i}\right) ,$ in a case study with six patterns
exposed to two different stimuli of the same intensity and duration (between
3000 and 4000 $n$ Monte Carlo trials). Each pattern is a string of $N$ bits.
Three patterns are randomly generated with 40, 50 and 60\% of the bits set to 1, and
the other three with the 1s at the first 70, 50 and 25\% positions,
respectively; the rest of the bits are set to $-1.$ The bottom graph shows
the baseline activity without stimulus (BS) and the activity level under
stimulus $\mu =1$ (SA1) and $\mu =2$ (SA2), i.e., two of the patterns. The
behavior which exhibits the system in this case (which we found for other
parameter values as well) is amazingly alike to observations in a comparable
(but true, not computer) experimental setting concerning the odor response
of the projection neurons in the locust antennal lobe \cite{Rabi,mazor}.

Interesting enough, the switching which shows our model due to stimulus
destabilization in the simulation of figure \ref{fig5} occurs for $\rho
<\rho _{c}.$ In fact, a similar phenomenon was observed also for $\rho
\rightarrow 0$ \cite{cortesNC}. This shows that, at least in this case, an
efficient adaptation to a changing environment does not require chaos.
However, the chaotic itinerancy we described above allows for a more
efficient search of the attractors space in a way that was believed to hold
in relevant systems under a critical condition \cite{chialvo}. Our
model thus illustrates a mechanism that makes chaos extremely beneficial.
This confirms expectations \cite{caos1,caos2,attent3} that the instability
inherent to chaos facilitates moving to any pattern at any time. The present
model system illustrates a specific mechanism which allows for this. As $%
\rho $ increases in a chaotic region, it is more likely that the activity
will visit all the attractors, not only the most correlated ones. The number
and diversity of attractors it visits then increases with $\rho ,$ and we
observed that the time spent in the attractor also varies with $\rho .$ The
system in this way may perform family discrimination and classification by
tuning $\rho .$ We finally remark that our model allows for describing a
coupling of $\rho $ to the activity, which may be quite a realistic setting
in some cases. No doubt it would be interesting to study other related model
situations.

We thank I. Erchova, P.L. Garrido and H.J. Kappen for very useful comments,
and financial support from FEDER--MEC project FIS2005-00791, JA project
P06--FQM--01505, and EPSRC--COLAMN project EP/CO 10841/1.

\newpage

\newpage
\section*{Figure Captions}

\noindent
{\bf Figure 1:}
The
Lyapunov exponent (solid curve), showing transitions from regular ($\protect%
\lambda <0$) to chaotic ($\protect\lambda >0$) as the \textit{%
synchronization parameter} $\protect\rho =n/N$ is varied, as obtained
analytically from the saddle--point solution for $\Phi =0.005,$ $M=1$
patterns, and $\protect\beta =50.$ The chaotic windows here were precisely
confirmed using related Monte Carlo simulations with $N=3600$ neurons. The minimum fraction of active neurons needed to start the period-doubling route to chaotic behavior, $\protect\rho _{c},$ is shown. 
This picture is strongly dependent on $\Phi ;$
there is a rather broad range of $\Phi $ values, including negative ones,
for which the behavior is qualitatively similar. The dashed curve is the
Hopfield--Hebb case $\Phi =1.$ The inset details the interesting region showing chaotic behavior.

\noindent
{\bf Figure 2:}
The overlap as a function of time (in units of $n$ Monte Carlo
trials) after $t=1920,$ for $N=1600,$ $\protect\beta =20,$ $\Phi =-0.4,$ $%
M=3 $ uncorrelated patterns and, from bottom to top, $\protect\rho =0.08,$
0.50, 0.65, 0.92 and 1.00, respectively. In this case, $\protect\rho %
_{c}=0.085.$

\noindent
{\bf Figure 3:}
Time variation of the mean firing rate $m\equiv(1+\pi)/2$ in an attractor neural network which stores a single pattern with depressing synapses, as modeled in \cite{depre0,depre1}, under partial updating in the oscillatory regime. Panels show, from top to bottom, the cases $\rho=1,0.7,0.3,0.1$. This (which corresponds to certain model parameters) reveals that, except for scaling of the typical temporal scale for the oscillations, partial updating does not introduce new phenomenology in this model, contrary to the case presented in this paper.

\noindent
{\bf Figure 4:}
This compares the gain function in the model in this paper, for $\rho=1$ and varying $\Phi$ (left panel) and the gain function in the model in \cite{storage} for varying $\gamma$ (right panel). In both cases $\beta$ was set to $3.$ Different curves in the left case are for 
$\Phi=1$ (non-depressed case), 0.6, 0.2 and 0 (hight depression); the curves in the right case occur when the corresponding parameter $\gamma=0 ~(\mbox{non-depressed case}),0.5,3,10 ~(\mbox{high depression case})$ 
This shows how the gain function can be non-monotonic for some values of the depression parameter $\Phi$ in the model in this paper. This allows for non-zero fixed point solutions, namely, the points that intersect the diagonal, with negative slopes (whose absolute value is larger than one) which leads to a period-doubling route to chaos.

\noindent
{\bf Figure 5:}
Itinerancy induced by external stimuli. Mean firing rates as a
function of time (bottom) and \textit{phase--space} trajectories (top)
trying to recreate an experimental observation concerning odor responses 
\protect\cite{mazor}. The graphs show two Monte Carlo simulations of our
system with $N=1600,$ $\protect\beta =4,$ $\Phi =-0.45,$ $\protect\rho =3/64<%
\protect\rho _{c},$ and six stored patterns, for different stimuli,
corresponding to green and red colors, respectively. The top graph involves
a standard false--neighbor method \protect\cite{embed} with \textit{%
embedding dimension }$d_{e}=5,$ and the time delay is $\protect\tau =20.$

\newpage

\end{document}